\documentclass[12pt]{iopart}
\usepackage{graphicx}

\usepackage{amsfonts}
\usepackage{amssymb} 

\begin{document}

\title[Hollow pellet injection for magnetic fusion]{Hollow pellet injection for magnetic fusion}

\author[LANL]{Zhehui Wang\textsuperscript{1}, M. A. Hoffbauer\textsuperscript{1}, E. M. Hollmann\textsuperscript{2}, Z. Sun\textsuperscript{3,4}, \\ Y. M. Wang\textsuperscript{3}, N. W. Eidietis\textsuperscript{5}, Jiansheng Hu\textsuperscript{3}, R. Maingi\textsuperscript{4}, \\ J. E. Menard\textsuperscript{4} \& X. Q. Xu\textsuperscript{6}}


\address{\textsuperscript{1} Los Alamos National Laboratory, Los Alamos, NM 87545, USA}
\address{\textsuperscript{2} University of California -- San Diego, La Jolla, CA 92093, USA}
\address{\textsuperscript{3} Institute of Plasma Physics, Chinese Academy of Sciences, Hefei, AH 230031, China}
\address{\textsuperscript{4} Princeton Plasma Physics Laboratory, Princeton, NJ 08544, USA}
\address{\textsuperscript{5} General Atomics, PO Box 85608, San Diego, CA 92186-5608, USA}
\address{\textsuperscript{6} Lawrence Livermore National Laboratory, Livermore, CA 94550, USA}

\ead{zwang@lanl.gov}
\vspace{10pt}
\begin{indented}
\item[] (Manuscript prepared for reviews by {\it Nuclear Fusion}, following the initial presentation in the 27th IAEA FEC, Gandhinagar, India, Oct. 22 - 27, 2018)
\end{indented}

\begin{abstract}
Precise delivery of mass to burning plasmas is a problem of growing interest in magnetic fusion. The answers to how much mass is necessary and sufficient can vary depending on parameters such as the type of atoms involved, the type of applications, plasma conditions, mass injector, and injection timing. Motivated by edge localized mode (ELM) control in H-mode plasmas, disruption mitigation and other applications in magnetic fusion, we report progress and new possibilities in mass delivery based on hollow pellets.  Here, a hollow pellet refers to a spherical shell mass structure with a hollow core. Based on an empirical model of pellet ablation, coupled with BOUT++ simulations of ELM triggering threshold, hollow pellets are found to be attractive in comparison with solid spheres for ELM control. By using hollow pellets, it is possible to tailor mass delivery to certain regions of edge plasmas while minimizing core contamination and reducing the total amount of mass needed. We also include experimental progress in mass delivery experiments, in-situ diagnostics and hollow pellet fabrication, and emphasize new experimental possibilities for ELM control based on hollow pellets. A related application is the disruption mitigation scheme using powder encapsulated inside hollow shells. Further experiments will also help to resolve known discrepancies between theoretical predictions and experiments in using mass injection for ELM control and lead to better predictive models for ELM stability and triggering. 
\end{abstract}

%
%
%
\maketitle
%
%

\section{Introduction}
One of the latest applications of mass injection in magnetic fusion is for Edge Localized Mode (ELM) control. ELMs are a key signature of high-confinement or H-mode plasmas. An H-mode mode plasma has a steep edge plasma pressure gradient and the spontaneous bootstrap current, or the `edge pedestal' which provides a transport barrier to improve particle and energy confinement. The same pressure gradient and current give rise to natural ELMs and other MHD instabilities. The amount of energy released by ELMs is proportional to the stored plasma energy and can exceed 10\% in the extreme cases. Natural ELMs in ITER and alike can potentially accelerate the plasma facing wall deterioration and ELM control is therefore necessary for ITER and future fusion reactors~\cite{Zinkle:2014,Federici:2001,Naujoks:2006,Wang:2007,Loarte:2014, Maingi:2014,Wang:2016}. Cryogenic hydrogen pellets~\cite{Lang:2004}, impurity pellets as well as  3D edge magnetic fields have experimentally shown to be feasible in inducing ELMs at a frequency higher than the natural ELM frequency of a few Hz and can reduce the peak energy flux onto the divertor and other plasma-facing surfaces. Experimental results from JET, AUG, and DIII-D, EAST and others have indicated that ELM suppression and triggering depend on the amount of mass injected. Cryogenic deuterium pellets and impurity pellets of lithium have been used experimentally. Only a sufficiently large size of lithium pellet ($> 600$ $\mu$m in diameter) is able to trigger ELMs with 100\% certainty in EAST~\cite{SunB:2018,Lunsford:2018}. 

A number of questions remain open from both the physics and the technology prospectives regarding ELM control using mass injection. From the physics viewpoint, the complex interactions of injected mass with the large gradients near the plasma edge make the modeling and quantitative predictions difficult. Although qualitative agreement has been achieved between experiments and modeling, quantitative prediction of mass ablation rate, pellet penetration depth are yet to be obtained in particular for impurity pellets. In some cases, the experimentally used pellets for ELM triggering are typically larger than model prediction. In other cases, the penetration distances of the pellets in experiments are less than what are given by modeling. From technology point of view, questions related to mass injection include the amount of mass needed, the timing of the mass injection with respect to the natural ELM cycle and location of the mass injection with respect to the magnetic flux surfaces and the edge pedestal, and the correlations between the injected mass properties with the induced ELM amplitude, and the size of ELM footprint on the divertor and other plasma-facing surfaces. Additional experiments, together with modeling and injection technology improvements will be needed to further advance the mass injection for ELM control, and in different plasma conditions. 

In this paper, we examine hollow pellet injection for ELM control in magnetic fusion. Here, a hollow pellet refers to a spherical shell structure surrounding a hollow core. Most of the pellets used in magnetic fusion, including cryogenic fueling and impurity pellets for ELM control, are objects topologically equivalent to a solid sphere. A shell structure surrounding impurity mass such as tungsten in the core has been demonstrated as the TESPEL diagnostic~\cite{ST:2012}. A shell structure enclosing boron powder has recently been demonstrated in DIII-D for disruption mitigation~\cite{Eidietis:2018,Hollmann:2018}. Compared with the existing  mass injection techniques, hollow pellet injection is attractive for ELM control in the following ways: ELMs can be triggered while substantial impurity contamination to the plasma core can be avoided. 
Initial demonstration of hollow pellet injection may use various existing injectors. In addition to ELM control, development of hollow pellets can enrich the pellet options for shell pellets and the applications in magnetic fusion include disruption mitigation, diagnostics, wall conditioning, helium ash removal, impurity transport, {\it etc}.

The rest of the paper is organized as follows: We first present a theoretical model for pellet interactions with an H-mode plasma. The model is characterized by a number of parameters for edge plasma condition, ablation, ELM triggering threshold. Then the model is applied to Li and B pellets with a solid core and with a hollow core. Core impurity contamination is then compared for different pellet sizes, shell thicknesses, and pellet injection velocities. Prototype layered structures of boron and polymers are being fabricated and characterized. Experimental progress for hollow pellet injection using the existing technologies for solid pellets are discussed. A dual-filter imaging technique, which belongs to a class of spectral imaging techniques that can be selectively tailored to at least two characteristic wavelengths, can be used for further development and applications of hollow pellet injection in magnetic fusion.

\section{ELM triggering \& ablation models}
The ELM triggering involves multiple physical processes. We may separate the triggering process into pellet ablation, neutral atom ionization, cold plasma propagation, plasma thermalization, MHD mode growth, and ELM crash, taking advantage of the separations in temporal scales involved in ELM triggering. We will focus on two types of impurity pellets, namely lithium and boron, partially motivated by ongoing research using these materials in different forms and their additional benefits to wall conditioning in high-temperature plasma devices.
\subsection{Ablation models}
\begin{figure}[htbp] 
  \centering
   \includegraphics[width=5.0in, angle=0]{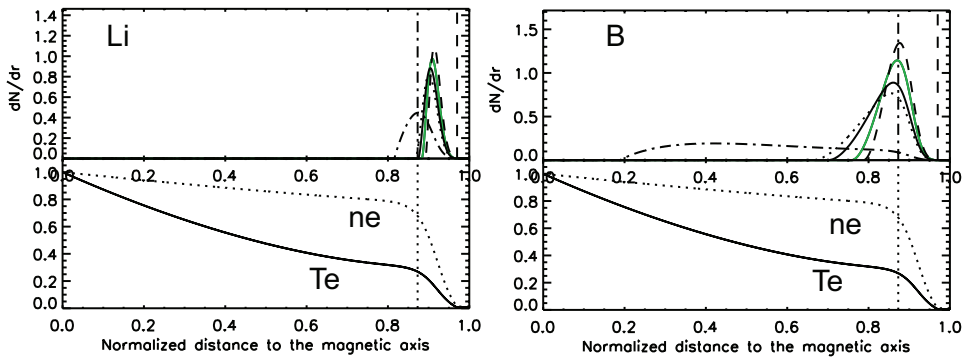}
      \caption{Comparison of various impurity ablation models. The left hand side is for lithium and the right hand side is for boron. $dN/dr$ is in the unit of 10$^{22}$ m$^{-1}$. The normalized electron density and temperature profiles (with respect to the peak temperature and density) are shown  in the frames below. Hollow pellets can achieve similar peak mass deposition while minimizing core plasma contamination. Pellet initial radius is 1 mm, initial velocity 100 m/s. The core plasma temperature is 5 keV, the core density is 8$\times$10$^{19}$ m$^{-3}$. The separatrix density is 10$^{18}$ m$^{-3}$ at a temperature of 30 eV. The pedestal width is 2\% of the minor radius.}
  \label{fig:abmodel1}
 \end{figure}

Plasma ablation of a pellet leads to the formation of a neutral cloud, which is subsequently ionized. The ionized atoms from a pellet propagate along the local magnetic flux tubes, justifying the assumption that mass deposition is local to a flux tube in a magnetized plasma. The amount of mass deposited onto the flux tube is given by
\begin{equation}
N_{ab} (r) = \int \frac{dN}{dt} dt  \sim <\frac{dN}{dt}> \frac{\Delta}{v_p},
\label{eq:abl}
\end{equation}
where $<dN/dt>$ is the average ablation rate, $\Delta$ the width of the flux tube and $v_p$ the pellet speed. $v_p$ is assumed to be a constant determined at the injector, which ignores the rocket effect. The average ablation rate $dN/dt$ (we remove the averaging `$<\cdot>$' symbol from now on for simplicity) can be in general written in the form~\cite{PLF:1988,SBKT:2006}
\begin{equation}
\frac{dN}{dt} = K n_{e0}^{k_1} T_{e0}^{k_2} r_p^{k_3},
\end{equation}
where the coefficient $K$ and the exponents $k_1$, $k_2$, $k_3$ for electron density $n_{e0}$ (prior to the pellet ablation), temperature $T_{e0}$ (prior to the pellet ablation) and pellet radius $r_p$ at the time of ablation depends on the microphysics of ablation as well as material properties such as the atomic number, ablation energy per atom, {\it etc}. Different empirical values for $K$ and $k_i$ ($i$ = 1 - 3) have been proposed~\cite{PLF:1988,SBKT:2006} and we show a few of them for an H-mode plasma in Fig.~\ref{fig:abmodel1}. 
%
%
%

Accordingly, in spite of their quantitative differences as indicated in Fig.~\ref{fig:abmodel1}, the predictions of mass deposition for different models are qualitatively similar. Below, we use one of the above models to examine the hollow pellet injection and compare it with solid pellet injection. Additional experiments using impurity pellets will help to constrain the ablation models better.

\subsection{ELM triggering mechanisms}
In the linear ideal MHD model, an ELM is triggered when the peeling-ballooning (P-B) mode becomes unstable and leads to a growth rate $\gamma > \omega_{*}/2$~\cite{Synder:2002}. In the nonlinear models and simulations~\cite{Xu:2010,Xi:2014}, the triggering threshold becomes $\gamma > \gamma_c$, with $\gamma_c \sim 0.1/\tau_A$ and $\tau_A$ being the Alfven time. Furthermore, JOREK simulations  also showed  that  ELM triggering  by  pellet  injection  is  correlated with the toroidally localized high edge pressure regions when the localized particle density increases due to the pellet ablation~\cite{Huysmans:2009,Futatani:2014}. Subsequent  increase in pressure is due to ionized pellet particle heating by the ambient plasma.  Recent simulations using BOUT++ came to similar conclusions~\cite{YMWang:2018}. When the pressure in this localized edge region exceeds  a  threshold,  ballooning  modes  grow  non-linearly leading  to  ELM  crash. In addition, the effects of the local electric field can not be ignored. In short, these works motivate an empirical density threshold scaling for ELM triggering ($N_{th}$) given by
\begin{equation}
N_{th} =\kappa_{th} a_0^{g_1} R_0^{g_2} \Delta^{g_3} n_{e0}^{g_4}, 
\label{eq:th1}
\end{equation}
here $\kappa _{th} > 1 $ is the threshold parameter for ELM triggering. $a_0$ is the minor radius, $R_0$ is the major radius, $\Delta$ the radial width of the ablation cloud, as given in Eq.~(\ref{eq:abl}), and $n_{e0}$ the local electron density before the pellet ablation. For simplicity, we choose $g_i=1$ ($i =$ 1 - 4). In this case, $\kappa_{th}$ is a dimensionless quantity that is a function of local normalized $\tilde{T}_e$, with the normalization energy being the averaged ablation energy per atom. Improvement of the scaling, which does not affect qualitatively the main conclusions here, will be deferred to further studies. For $R_0$ = 1.67 m and $a_0 =$ 0.6 m, $n_{e0}$ = 5$\times$10$^{18}$ m$^{-3}$, $\kappa_{th}$ is found to be 10 to 100 times the local plasma density. Correspondingly, $N_{th}$ = 1.7$\times$10$^{18}$ or the same number of atoms in a lithium solid sphere with a radius around 200 $\mu$m. For higher local plasma density, a larger pellet would be needed according to Eq.~(\ref{eq:th1}).

 


\section{Mass delivery and core impurity}
In an ITER type-I ELMy H-mode plasma, electron density and temperature are 4$\times$ 10$^{19}$ m$^{-3}$ and 500 eV respectively at the separatrix, and 8.7 $\times$10$^{19}$ m$^{-3}$ and 4 keV respectively at the top of the transport barrier. For comparisons of hollow and solid pellets, we use an H-mode profile that may be achieved in the existing devices to examine ELM triggering and theoretical predictions. Li and B pellets are compared here. The theoretical framework introduced can be readily extended to ELM triggering scenarios using other types of pellet materials and different plasma conditions.

\begin{figure}[htbp] 
  \centering
   \includegraphics[width=5.0in, angle=0]{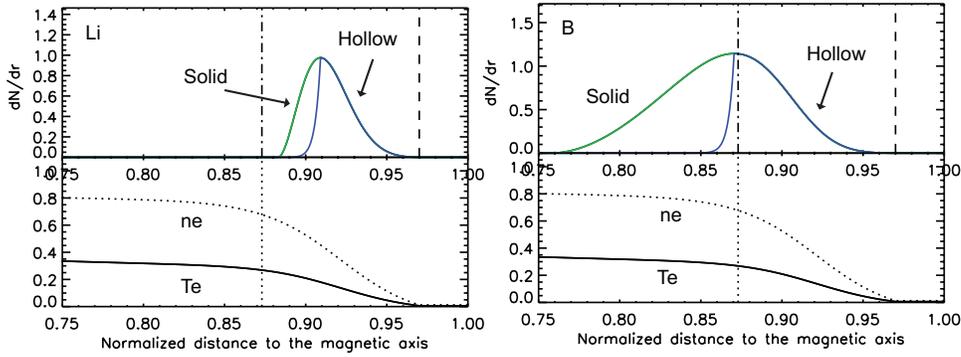} 
   \caption{Comparison of mass deposition of a solid-core pellet and a hollow pellet. The left hand side is for lithium and the right hand side is for boron. $dN/dr$ is in the unit of 10$^{22}$ m$^{-1}$. The normalized electron density and temperature profiles (with respect to the peak temperature and density) are shown  in the frames below. Hollow pellets can achieve similar peak mass deposition while minimizing core plasma contamination. Pellet initial radius is 1 mm, initial velocity 100 m/s. The peak plasma temperature at the core is 5 keV, the core density is 8$\times$10$^{19}$ m$^{-3}$. The separatrix density is 10$^{18}$ m$^{-3}$ at a temperature of 30 eV. The pedestal width is 2\% of the minor radius.}
  \label{fig:pp1}
 \end{figure}

In Fig.~\ref{fig:pp1}, the mass deposition as a function of normalized distance is shown for lithium and boron pellets with an initial radius of 1 mm and injection velocity of 100 m/s. Two vertical dashed lines in the top frames mark the boundaries of the pedestal region as in Fig.~\ref{fig:abmodel1}. In contrast with solid pellets, a sharper decrease in mass deposition after the ablation rate reaching its maximum is expected for both Li and B hollow pellets. For solid pellets, the deeper deposition of boron inside the pedestal top is due to the combination of two factors: its higher ablation energy per atom and larger number of atoms for a fixed size (about 3$\times$ larger). The average energy of ablation is 5.6 $\pm$ 0.6 eV per boron atom and 1.6 $\pm$ 0.4 eV per lithium atom. The strong dependence of the mass deposition on the ablation energy per atom is also confirmed with cryogenic hydrogen pellet of the same size (ablation energy 5.3 meV per atom), which reaches its peak of mass deposition at a shallower depth than both the lithium and boron pellets. 
 
The results in Fig.~\ref{fig:pp1} are consistent with earlier findings that, in order to reliably trigger an ELM, the pellet needs to be sufficiently large (and fast) to penetrate close to the pedestal top~\cite{Kocsis:2007, Lang:2011,Huijsmans:2015}. Although both solid and hollow pellets can trigger ELMs, the fact that ELM triggering is near the top of the pedestal potentially poses an issue of core contamination using impurity pellets as implied by Fig.~\ref{fig:pp1}. Strong atomic number ($Z$) -dependence of the core contamination may be expected when using impurity pellets for ELM control. 

\subsection{Core impurities}
\begin{figure}[htbp] 
  \centering
   \includegraphics[width=3.0in, angle=0]{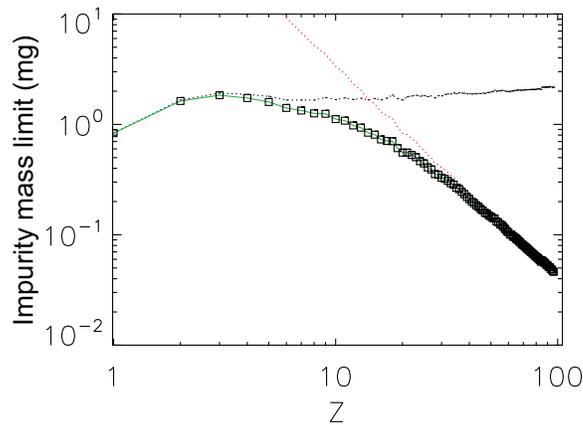} 
   \caption{The mass limits (in mg) for different elements using impurity fraction limits bound by dilution and radiation (`$\Box$' symbol). The two dashed lines (color coded online) correspond to the dilution and radiation limit respectively. The mass limit for lithium is 1.84 mg (0.94 mm in solid sphere radius), for boron is 1.60 mg (0.55 mm in solid sphere radius). The total number of plasma electron is assumed to be 1.0$\times$10$^{22}$.}
  \label{fig:imm1}
 \end{figure}

It is necessary to minimize the core impurity buildup when using mass injection or other methods for ELM pacing and control~\cite{Loarte:2014}. The impurities can come from a number of sources: the impurity pellets themselves, impurities released from the wall and divertor due to the heat and particle fluxes from ELMs, and the DT fusion generated helium. The impurity mass limits have been estimated as function of the atomic number ($Z$) using the known processes of dilution for low-Z and radiation for high-Z impurities~\cite{Cecil:1980}. A quantitative result using a similar empirical formula is given in Fig.~\ref{fig:imm1} for the total number of plasma electrons of 1.0$\times$10$^{22}$. For sufficiently small Z, the fraction of the atoms due to fuel dilution is limited to 0.05/Z of the total number of fuel ions or electrons. For example, the hydrogen concentration should not exceed 5\% of the electron density in the plasma. Helium fraction, including contributions from the fusion process, should be kept below 3\%. For high-Z atoms such as tungsten, the concentration is limited to 10/Z$^3$ or 2.5$\times$10$^{-5}$. Each ELM event is known to expel impurities from the plasma and to mitigate the buildup of plasma core impurities, although how far inside the separatrix the impurities are expelled during the ELMs needs further studies. The potential advantages of a hollow pellet that can trigger ELMs while minimizing the core impurity deposition warrant further experimental validation.
 
\subsection{Velocity and size dependence}
\begin{figure}[htbp] 
  \centering
   \includegraphics[width=3.5in, angle=0]{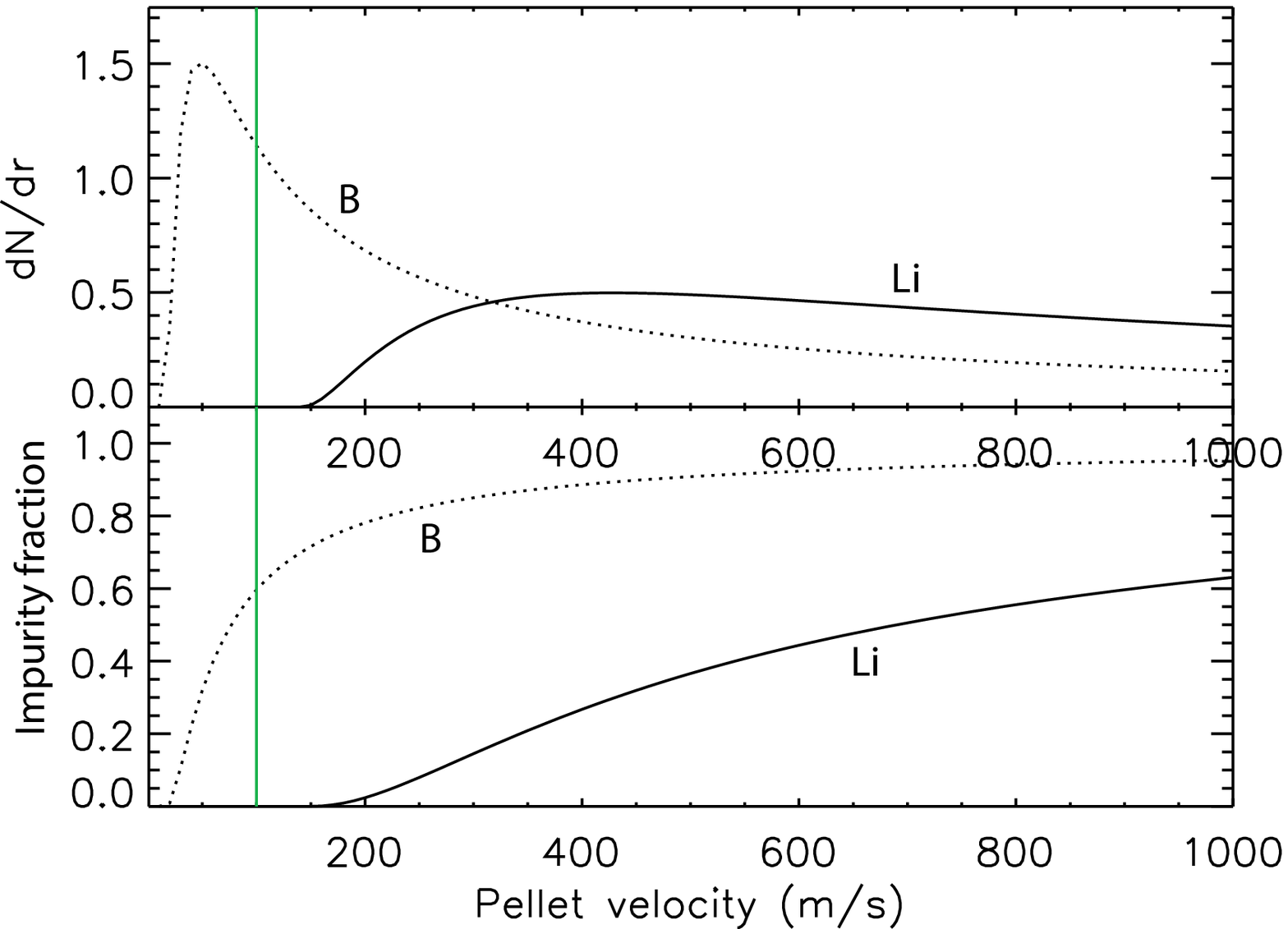} 
   \caption{For fixed pellet radius $r_p = 1$ mm and fixed pedestal conditions, (Top) the mass deposition ($dN/dr$ in 10$^{22}$ m$^{-1}$) as a function of Li or B solid sphere injection velocity. (Bottom) The fraction of the solid spheres that reach beyond the ELM triggering location as a function of injection velocity. The colored vertical line corresponds to the example shown in Fig.~\ref{fig:pp1}.}  
  \label{fig:fv1}
 \end{figure}

We further examine the mass deposition from solid pellets as a function of initial size and injection velocity. For simplicity, we shall consider pellet penetration at the constant initial injection velocity and ignore acceleration due to ablation and other plasma-pellet interactions. We also assume that the ELM triggering happens at a fixed minor radius. Specifically, at the location near the shoulder of pedestal top, or the normalized minor radius at 0.87 as shown in Fig.~\ref{fig:pp1}. The corresponding local electron temperature and density are 1.3 keV and 6.3$\times$10$^{19}$ m$^{-3}$. For the fixed pellet radius of 1 mm, the mass deposition as a function of pellet injection velocity is shown in Fig.~\ref{fig:fv1}.

\begin{figure}[htbp] 
  \centering
   \includegraphics[width=3.5in, angle=0]{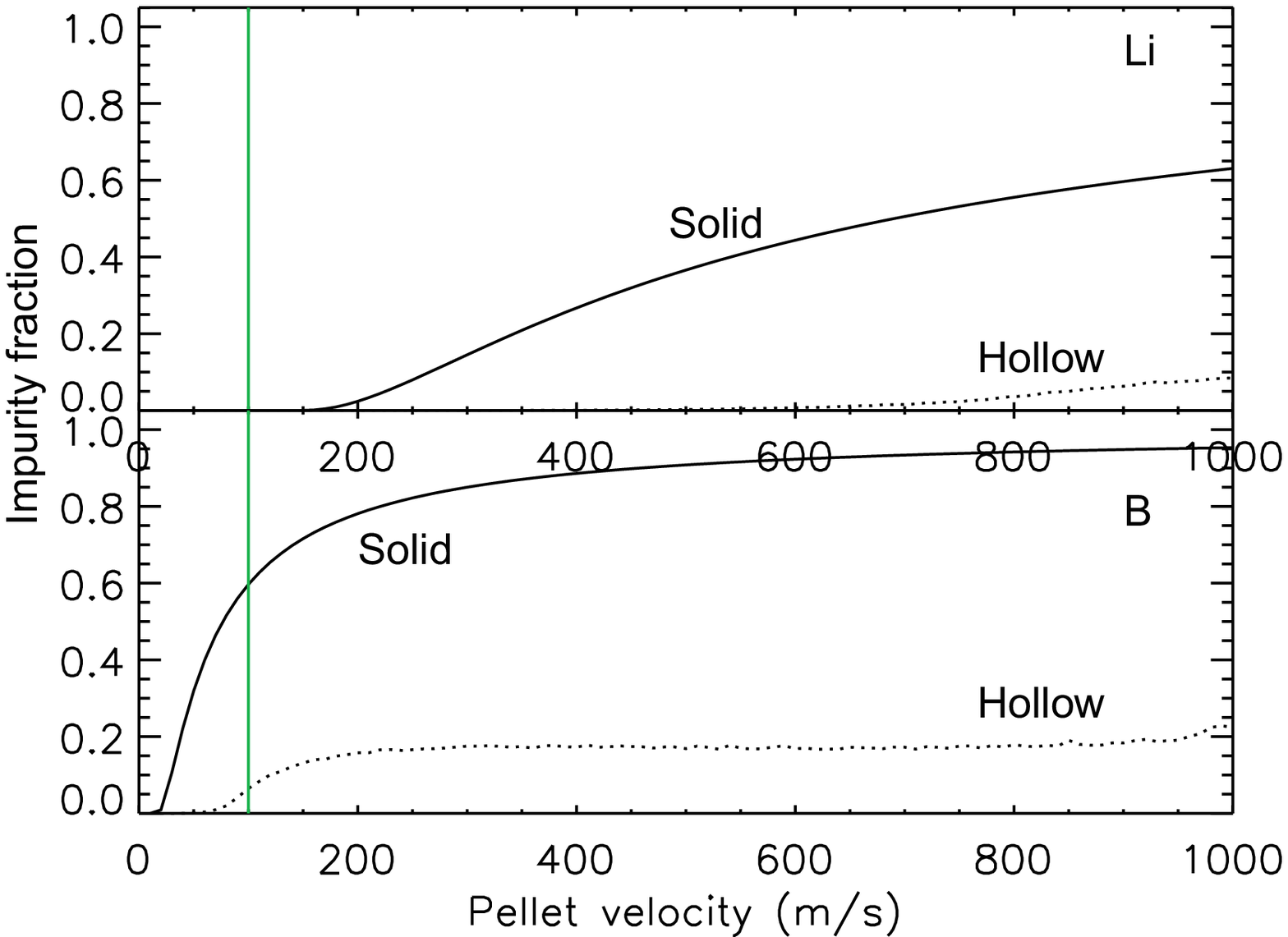} 
   \caption{For fixed pellet radius $r_p = 1$ mm and fixed pedestal conditions, (Top) the mass deposition ($dN/dr$ in 10$^{22}$ m$^{-1}$) as a function of Li or B solid sphere injection velocity. (Bottom) The fraction of the solid spheres that reach beyond the ELM triggering location as a function of injection velocity. The colored vertical line corresponds to the example shown in Fig.~\ref{fig:pp1}.}
  \label{fig:fv2}
 \end{figure}

The trends in mass deposition are similar for Li and B. A peak deposition is reached near 50 m/s for B and 450 m/s for Li. A significant amount of mass deposition is expected following the ablation at the targeted radius of 0.87 for high injection speeds and the amount increases with injection velocities, as shown in the lower frame in Fig.~\ref{fig:fv1}. Reduction of the impurity using hollow pellets of the same initial radius of 1 mm is shown in Fig.~\ref{fig:fv2}. Furthermore, the observation that a relatively flat or plateau region between 200 to 800 m/s exists for B implies that hollow pellet may also relax the requirements on the precise injection velocity for impurity control.

 A number of existing impurity launchers may be used to achieve the injection speeds as required in Fig.~\ref{fig:fv1} and Fig.~\ref{fig:fv2}~\cite{Wang:2016}. Different launchers may be distinguished by their different forces of acceleration. We would like to mention that the mechanical strength of pellets may limit the maximum force of acceleration, in particular for thin hollow pellets~\cite{WK:2003}. A remedy is to increase the launcher size to accommodate a long pathlength of acceleration. Furthermore, thin shell pellet may open doors to alternative acceleration methods such as electrostatic acceleration that is currently not in use for magnetic fusion.
 
\begin{figure}[htbp] 
  \centering
   \includegraphics[width=3.5in, angle=0]{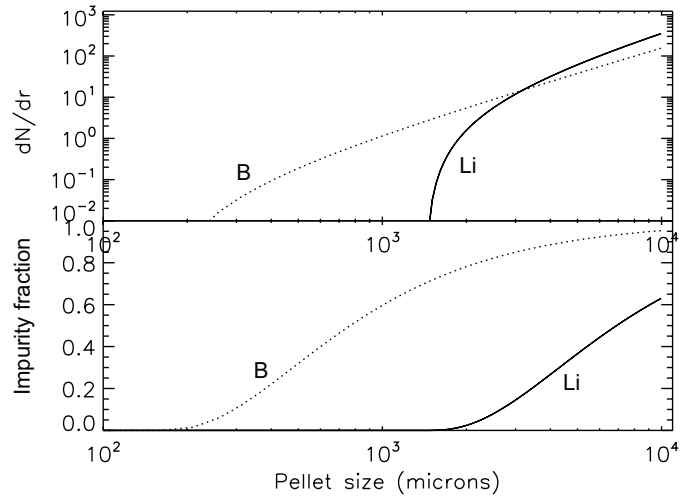} 
   \caption{For fixed pellet velocity $V_p = 100$ m/s and fixed pedestal conditions, (Top) the mass deposition ($dN/dr$ in 10$^{22}$ m$^{-1}$) as a function of Li or B solid sphere size (radius). (Bottom) The fraction of the solid spheres that reach beyond the targeted ELM triggering location as a function of solid sphere size.}
  \label{fig:ss1}
 \end{figure}

\begin{figure}[htbp] 
  \centering
   \includegraphics[width=3.5in, angle=0]{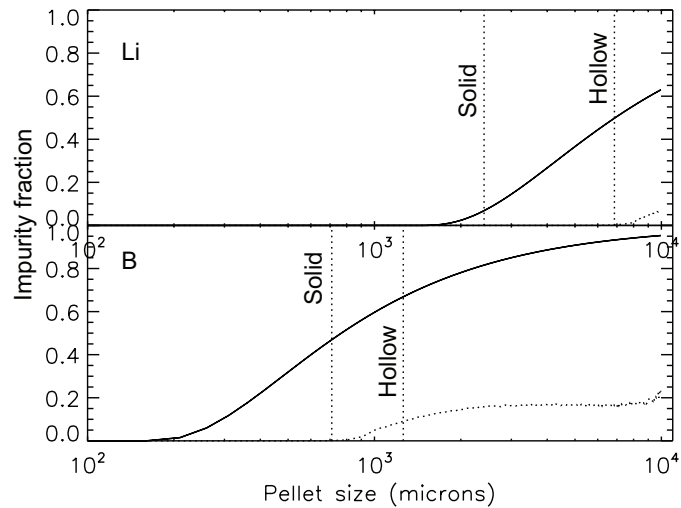} 
   \caption{For fixed pellet velocity $V_p = 100$ m/s and fixed pedestal conditions, core impurity reduction as a function of size by using hollow pellets. (Top) Lithium impurity fraction beyond the targeted ELM triggering location; (Bottom) Boron impurity fraction beyond the targeted ELM triggering location beyond the targeted ELM triggering location. The vertical dashed lines corresponds to impurity limits given in Fig.~\ref{fig:imm1}.}
  \label{fig:redss1}
 \end{figure}

  For the fixed pellet injection velocity of 1 mm, the mass deposition as a function of pellet radius is shown in Fig.~\ref{fig:ss1}. The mass deposition increases with the pellet size as expected. Meanwhile, the impurity fraction that can contaminate the plasma core also increases. In Fig.~\ref{fig:redss1}, the vertical lines are now corresponds to the limits set by Fig.~\ref{fig:imm1} for boron and lithium. Hollow pellet structure allows a wider selection of radius than solid structure within the impurity limits.

\subsection{Optimal hollow pellet dimensions}
\begin{figure}[htbp] 
  \centering
   \includegraphics[width=3.5in, angle=0]{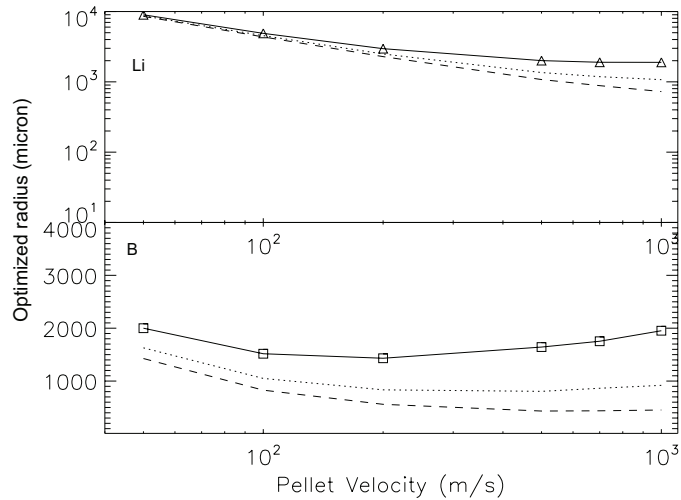} 
   \caption{Optimal hollow pellet initial radius depends on the atomic number, injection velocity, pedestal condition. The calculated Li and B radius as a function of injection velocity for fixed pedestal conditions are shown here for different amount of material needed at the pedestal shoulder (10$^{19}$ -- dashed lines, 10$^{20}$ -- dotted lines and 10$^{21}$ -- solid lines). The results also indicate that the injection can be further optimized by tuning the injection velocity.}
  \label{fig:pv1}
 \end{figure}

Based on the above, the dimensions of an optimal spherical hollow pellet at a fixed injection velocity $v_p$, {\it i.e.}, its initial radius and thickness, are determined by the following factors: the pedestal plasma condition, the amount of mass required to trigger an ELM, the tolerable impurity levels, and the atomic number. As mentioned above, material strength also needs to be considered for hollow pellet injection, which may limit the launcher selections and acceleration methods when high injection speed is required at a short acceleration pathlength. The total hollow pellet mass is thus given by
\begin{equation}
M_h = M_1 + M_2 + M_3,
\end{equation}
where $M_1 = N_{th} m_0$ is determined by the ELM triggering threshold, with $N_{th}$ given by Eq.~(\ref{eq:th1}) and $m_0$ the atomic mass of the pellet. Pellets of chemical compounds can use the molecular mass instead of $m_0$ for an atomic mass. $M_2$ is the mass loss before the pellet reaches the targeted ELM triggering location, which is partially determined by the pedestal plasma condition. $M_3$ is the residual pellet mass passing through the targeted ELM triggering location, which is limited by the impurity tolerance level and ideally $M_3 =$ 0. In the case studies as in Fig.~\ref{fig:pp1}, $M_2$ sets the lower bound in the hollow pellet shell thicknesses ($T_h$), which are found to be 158 $\mu$m (B) and 1424 $\mu$m (Li) respectively for $v_p =$ 100 m/s. At $v_p = 400$ m/s, one finds the minimal boron shell thickness to be 59 $\mu$m and lithium shell thickness to be 356 $\mu$m. For a fixed $N_{th}$ = 10$^{19}$, 10$^{20}$ and 10$^{21}$, The optimized pellet radius is summarized for Li and boron in Fig. ~\ref{fig:pv1}.

According to Eq.~(\ref{eq:abl}), $M_1$ corresponds to
\begin{equation}
\frac{M_1}{m_0} = \kappa_2 (r_p-T_h) \frac{dN}{dr}.
\end{equation}
Here $(r_p-T_h)$ is the instantaneous pellet radius at the ELM triggering location. $T_h$ represents the pre-triggering shell loss. Previous studies indicate that the width of the mass deposition can be multiple times the instantaneous pellet radius, with the multiplier ($\kappa_2$) being in the range of 2 to 10. We use $\kappa_2 =$ 5 in Fig.~\ref{fig:pv1}. The smallest pellet that can satisfy all the condition is therefore given by
\begin{equation}
r_p^{min} = \frac{M_1}{\kappa_2 m_0 dN/dr}+T_h
\end{equation}

\section{Hollow pellet fabrication}
We  briefly go through some existing options before summarizing the progress in developing hollow boron spheres.
\subsection{Existing options}
\begin{figure}[htbp] 
  \centering
   \includegraphics[width=4.0in, angle=0]{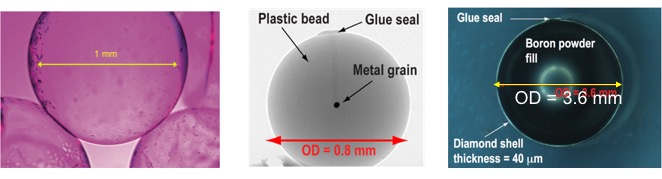} 
   \caption{Examples of various core-shell pellets. (a.) Hollow polymer pellet developed for ICF applications; the shell thickness is about 20 $\mu$m. (b.) A plastic bead impregnated with a small tungsten grain, this is used as a calibration pellet in DIII-D; (c.) A diamond shell filled with boron powder. This is developed for disruption mitigation in DIII-D. The diamond shell thickness is 40 $\mu$m.}
  \label{fig:hs1}
 \end{figure}

Various hollow spherical targets have been developed for inertial confinement fusion (ICF) experiments~\cite{Hoppe:2000,Nagai:2004,Du:2016}. Examples include hollow glass spheres, hollow polymer spheres (an example is shown in Fig.~\ref{fig:hs1}), hollow boron carbide spheres. While some of these targets may be directly used to magnetic fusion (MF) applications, broadening the material and structure choices are of  interest for a number of reasons. The ICF targets are designed for DT fusion under extremely high pressure for a brief time window, which is estimated to be around 10$^{-11}$ s. 

Core-shell spheres have also been developed and adopted in MF applications, see two examples shown in Fig.~\ref{fig:hs1}. Small hollow pellets are excellent tools for calibration of spectroscopic diagnostics in tokamaks, as a well-known quantity of the desired material can be delivered to the plasma core~\cite{ST:2012}. The shell protects the core material from ablation and loss in the launch tube and plasma edge region and ensures that the desired quantity of material to be studied reaches the core. In the DIII-D experiments~\cite{Hollmann:2017}, small (OD = 0.8 mm) plastic (poly-alpha methylstyrene, PAMS, C$_9$H$_{10}$) plastic bead pellets carrying much smaller (10 $\mu$g) tungsten grains were used to calibrate spectroscopic core tungsten measurements in support of the DIII-D tungsten divertor rings experiments. Another type of core-shell also been pursued on the DIII-D tokamak for disruption mitigation studies~\cite{Eidietis:2018,Hollmann:2018}.

We would like to point out some differences and commonalities between the traditional ICF applications and the proposed new MF applications. Deployment of an ICF target does not involve significant motion or acceleration. Motion and acceleration could put additional requirement in the structural rigidity. The key requirements in ICF hollow targets are sphericity and uniform wall thickness of the shells. Impurity control is important for magnetic fusion applications as mentioned above. Another unique feature of MF requirement is that the hollow sphere size may be larger. An ICF target is estimated to 1-2 mm in radius, constrained by the driver laser power and fusion power gain considerations.

\subsection{Boron spheres}
\begin{figure}[htbp] 
  \centering
   \includegraphics[width=4.0in, angle=0]{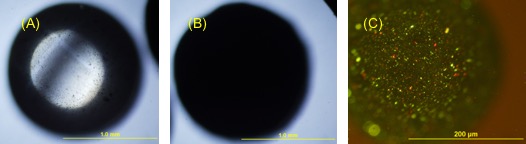} 
   \caption{Images of boron-shell enclosed PMMA spheres. (a.) A PMMA sphere that is used as a template for boron coating; the sphere is  1.5 $\pm$ 0.05 mm in diameter and a sphericity of 50 $\mu$m. (b.) A boron coated PMMA sphere; (c.) The same sphere as in (b) with a higher microscope magnification.}
  \label{fig:fB1}
 \end{figure}

A growing number of methods are being developed to make spheres and hollow spheres, for example, the microfluidic techniques~\cite{Nisisako:2016}.  We report progress with hollow boron shell development, using an approach similar to a gel-casting method described recently to make boron carbide (B$_4$C) hollow spheres~\cite{Chen:2017}.  The fabrication took two main steps. In the first step, core-shell structured B$_4$C was fabricated by coating molybdenum (Mo) balls with B$_4$C slurry. The fabricated core-shell structured B$_4$C microspheres exhibit a large size (2200 -- 2300 $\mu$m) and a wall thickness of 100 -- 180 $\mu$m. In the second step, the core-shell structured B$_4$C microspheres were laser drilled and the metal cores subsequent corrosion to obtain the B$_4$C hollow microspheres. It should be mentioned that the gel-casting technique is also suitable for the preparation of other ceramic hollow microsphere that may be of interest to magnetic fusion, including assessment of material and first wall options. We have made the first samples of boron shell with a PMMA core, as shown in Fig.~\ref{fig:fB1}. 

\section{Experimental progress}
\begin{figure}[htbp] 
  \centering
   \includegraphics[width=5.0in, angle=0]{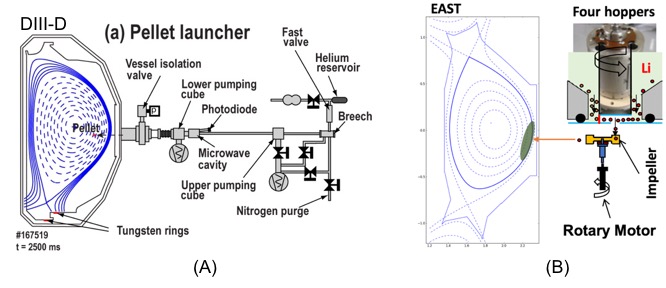} 
   \caption{(A) Schematic of a pellet launcher used for various pellet injection in DIII-D. (B) Schematic of a four-hopper Li granule injector deployed on the EAST. This apparatus is able to inject four different sizes of pellets in the range of 0.2 to 1 mm. The velocity ranges from 30 to 110 m/s for Li. }
  \label{fig:EH1}
 \end{figure}

Several existing techniques can be leveraged upon towards demonstration of the new hollow pellets in MF. One is a pellet launching system. Another is a diagnostic system that can monitor the pellet-plasma interactions in real-time with good spatial and temporal resolutions. As the first example, 
a pneumatic (helium pulse) launcher in DIII-D can inject pellets radially inward at velocities of 100 -- 300 m/s from the outer midplane~\cite{Hollmann:2017}. A schematic of the experimental setup is shown in Fig.~\ref{fig:EH1}. The pellet launcher breech is flexibly configurable to allow holding and firing of different sized pellets.
 
 

\begin{figure}[htbp] 
  \centering
   \includegraphics[width=2.7in, angle=0]{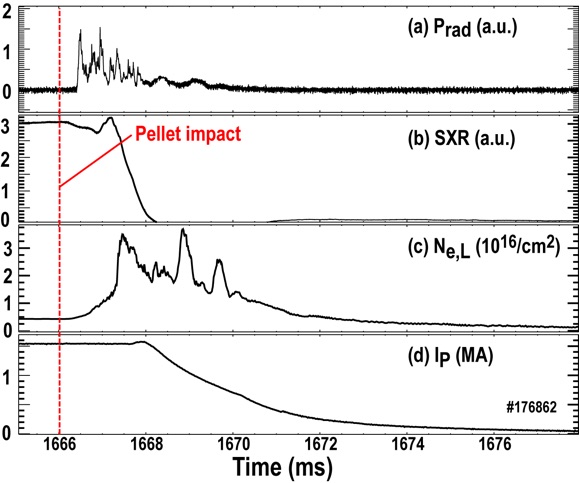} 
   \caption{Example of rapid shutdown with large (OD = 3.6 mm) shell pellet showing rapid radiative shutdown. Time traces are shown of (a) radiated power, (b) SXR brightness (showing collapse of thermal energy), (c) electron line density (showing deposited impurities) and (d) plasma current (showing decay of magnetic energy).}
  \label{fig:d3d1}
 \end{figure}

Recently, the pneumatic launcher system was first successfully used to demonstrate the shell pellet concept for disruption mitigation in DIII-D~\cite{Eidietis:2018, Hollmann:2018}. A picture of the pellet was shown in Fig.~\ref{fig:hs1}, which has an OD of 3.6 mm made of diamond with a shell thickness of 40 $\mu$m. The shell is filled with boron powder. The pellet shells burnt through close to the plasma magnetic axis, releasing boron powder and causing a very rapid radiative shutdown of the plasma, as shown in Fig.~\ref{fig:d3d1}. Large hollow pellets (`shell pellets') filled with different payloads are of interest for tokamak disruption mitigation, since a precisely designed payload can be delivered to the plasma core and, ideally, satisfy the different shutdown requirements to minimize different times of wall damage that can result from disruptions, such as localized heat loads and vessel forces. Previous shell pellet disruption mitigation experiments have had challenges getting payload deposition into the core during the disruption, as the pellets have either passed completely through the plasma without breaking open or have broken in the plasma edge or have not caused a rapid shutdown~\cite{Hollmann:2010}. Future work will continue to study the use of shell pellets for disruption mitigation in DIII-D for application in future large tokamaks like ITER and DEMO.
 
\begin{figure}[htbp] 
  \centering
   \includegraphics[width=2.7in, angle=0]{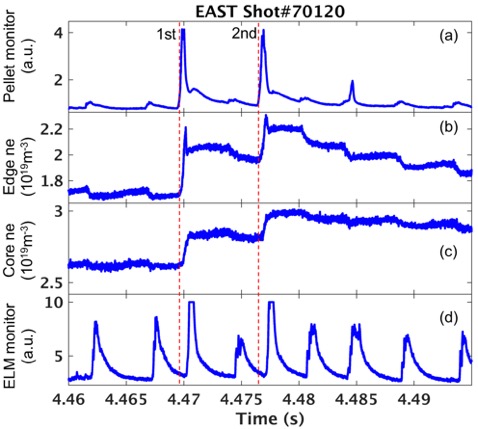} 
   \caption{Typical two Li granules injection for ELM trigger on EAST. Time traces are shown of (a)granule injection monitor, integrated density from edge chord and (c)core chord, and (d) D$_\alpha$ pointing to divertor for ELM monitor.}
  \label{fig:east1}
 \end{figure}

 EAST recently deployed a multi-chamber device to inject four different Li pellets for ELM pacing, see Fig.~\ref{fig:EH1}B, aiming at decoupling ELM triggering from the fueling effects of deuterium pellets~\cite{SunB:2018}. By horizontally injecting Li pellets with velocities around 80 m/s, diameters ranging from 200 microns to 1 mm into the low field side of EAST H-mode discharges, it was demonstrated that Li pellets with diameters above 600 microns can successfully trigger ELMs more than 95\% of the time~\cite{Lunsford:2018}. When using 600 micron and larger solid lithium pellets ($\sim$ 5.24 $\times$ 10$^{18}$ atoms) however, fueling of electron density was observed. Fig.~\ref{fig:east1} shows the typical 0.6 mm lithium pellets are injected into H-mode plasma. The H-mode discharge with plasma current 400 kA was sustained by about 6 MW auxiliary heating power.  The average density is about 2.7$\times$10$^{19}$ m$^{-3}$, maintained by feedback control via supersonic molecular beam injection. The two-pellet injection scheme produced sufficient edge plasma perturbation and triggered two ELMs, which was confirmed by the D$_\alpha$ spikes and edge density crash just following the pellet injection (the dashed lines in Fig.~\ref{fig:east1}). Although the ELM crash caused the edge density reduction from 2.2 to 2 , the edge density and core density were still $\sim$ 20\% higher than prior to the pellet injections. After the second pellet injection, the edge and core density rised further. The edge and core density reduced gradually by spontaneous ELMs. The core density reduction was slower than that of the edge density. These observations indicate that the pellet-trigged ELM cannot expel enough particles to maintain constant density in the plasma core. A large solid Li pellet, which is sufficient to trigger ELMs, cannot alleviate the fueling effect to the plasma core. Furthermore, increased radiation has also been observed correlated with Li pellet injection. The likely explanation is that the impurity ions from the pellet core, which likely contribute little to ELM triggering, but definitely to the electron density increase in the plasma core. Hollow pellets offer a promising method to solve this and related issues.
 
For the in-situ diagnostic of a hollow-pellet-plasma interaction, we may use a recently demonstrated dual-filter imaging technique~\cite{Sun:2018}, the structures of pellet ablation is resolved using the new technique. The imaging technique can be used in conjunction with a hollow-pellet injection experiment. 

\begin{figure}[htbp] 
  \centering
   \includegraphics[width=2.7in, angle=0]{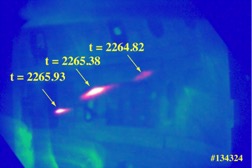} 
   \caption{False color images (C-II emission) of spherical plastic calibration pellet moving through DIII-D plasma~\cite{Hollman:2009}.}
  \label{fig3:d3d1}
 \end{figure}

\newpage
\section{Summary \& conclusions}
Precise delivery of mass to magnetic fusion plasmas is a problem of growing interest. We used various empirical models for ablation, augmented by BOUT++ simulations, to assess the concept of hollow pellets for ELM applications in magnetic fusion. Hollow sphere for precise ELM control is possible with significant reduction of impurity contamination in the core. Prototype core-shell boron spheres have been fabricated. Additional applications of hollow spheres in diagnostics and fueling are also possible. 

{\it Acknowledgement} We thank Dr. Peter Goodwin (CINT/LANL) for help with microscopy of the boron spheres. LANL work is supported in part by the US Department of Energy (DoE) Fusion Energy Sciences long-pulse tokamak program through the Triad National Security, LLC (`Triad') contract \# 89233218CNA000001. PPPL work is supported by US DoE through the contract \# DE-AC02-09CH11466. DIII-D material is based upon work supported by US DoE under Award Number DE-FC02-04ER54698.

{\it Disclaimer} This report was prepared as an account of work sponsored by an agency of the United States Government.  Neither the United States Government nor any agency thereof, nor any of their employees, makes any warranty, express or implied, or assumes any legal liability or responsibility for the accuracy, completeness, or usefulness of any information, apparatus, product, or process disclosed, or represents that its use would not infringe privately owned rights.  Reference herein to any specific commercial product, process, or service by trade name, trademark, manufacturer, or otherwise, does not necessarily constitute or imply its endorsement, recommendation, or favoring by the United States Government or any agency thereof.  The views and opinions of authors expressed herein do not necessarily state or reflect those of the United States Government or any agency thereof.

\end{document}